\documentclass{article}

\PassOptionsToPackage{numbers, compress}{natbib}

\usepackage[preprint]{neurips_2026}

\usepackage[utf8]{inputenc} 
\usepackage[T1]{fontenc}    
\usepackage{hyperref}       
\usepackage{url}            
\usepackage{booktabs}       
\usepackage{amsfonts}       
\usepackage{nicefrac}       
\usepackage{microtype}      
\usepackage{xcolor}         
\usepackage{amsmath}
\usepackage{multirow}
\usepackage{graphicx}
\usepackage[table]{xcolor}
\definecolor{mygray}{gray}{0.9}
\usepackage{enumitem}
\usepackage{wrapfig}
\usepackage{subcaption}

\title{Direct Preference Density Alignment for Conversational Audio Equalization}

\author{%
  Ioannis Stylianou\\
  Department of Electronic Systems\\
  Aalborg University\\
  Aalborg, Denmark \\
  \texttt{iost@es.aau.dk} \\
  \And
  Sven Ewan Shepstone \\
  Bang \& Olufsen A/S \\
  Struer, Denmark \\
  \texttt{ssh@bang-olufsen.dk} \\
  \And
  Jon Francombe \\
  Bang \& Olufsen A/S \\
  Struer, Denmark \\
  \texttt{jofr@bang-olufsen.dk} \\
  \And
  Pablo Martinez Nuevo \\
  Bang \& Olufsen A/S \\
  Struer, Denmark \\
  \texttt{pmn@bang-olufsen.dk} \\
  \And
  Zheng-Hua Tan \\
  Department of Electronic Systems\\
  Aalborg University\\
  Aalborg, Denmark \\
  \texttt{zt@es.aau.dk} \\
}

\begin{document}

\maketitle

\begin{abstract}
Large Language Model alignment typically relies on learned proxy reward models, which significantly increase the memory footprint during training and are notoriously prone to instability and reward hacking. While offline methods like Direct Preference Optimization (DPO) bypass the reward model, they lose the ability to perform online exploration. If no optimization constraints are applied, this can lead to format collapse in bounded, continuous spaces. To resolve this, we propose Direct Preference Density Alignment: An alternative framework that removes the need for a learned proxy reward model while strictly preserving the benefits of online reinforcement learning. We leverage large-scale user data (approximately 90,000 samples) to construct non-parametric preference density maps, establishing an empirical reward surface. In addition to removing the reward model, Direct Preference Density Alignment enables the combination of the online structural grounding of Group Relative Policy Optimization (GRPO) with the targeted offline refinement of DPO. We show that this GRPO+DPO combination achieves the highest performance, and in a blind audio equalization listening test, enables a 1.5B-parameter model to achieve perceptual parity with a carefully prompt-engineered GPT-4o mini baseline, using only a fraction of the inference compute.
\end{abstract}

\section{Introduction}
\label{sec:introduction}
Aligning Large Language Models (LLMs) to subjective continuous-control tasks remains a significant open challenge. Modern Reinforcement Learning from Human Feedback (RLHF) \cite{ouyang2022training} pipelines rely heavily on Proximal Policy Optimization (PPO) and learned proxy Reward Models (RMs). However, RMs are highly susceptible to ``reward hacking'' \cite{gao2023scaling} and PPO is extremely resource-intensive, requiring four separate LLMs (Trained, Reference, Reward, and Value models) during training.

Recent advancements attempt to mitigate this compute burden. Group Relative Policy Optimization (GRPO) \cite{shao2024deepseekmath} eliminates the value model, while offline methods like Direct Preference Optimization (DPO) \cite{rafailov2023direct} eliminate both the value and reward models. However, we demonstrate that by shifting from online Reinforcement Learning (RL) to offline ``contrastive'' learning, DPO loses active exploration. In compact spaces \cite{frechet1906quelques}, this lack of environmental (or programmatic) grounding leads to format collapse, where the model forgets valid syntactical constraints.

To resolve this, we propose a proxy-free, online RL framework for continuous parameter estimation. Instead of training an unstable proxy RM, we leverage population data to construct static, non-parametric preference surfaces. The models then navigate the surface like a topographical map, exploring different regions to eventually find the ``peaks'' of highest human satisfaction. We instantiate this framework on a subjective audio equalization task, mapping abstract natural language (e.g., ``make it warmer'') to loudspeaker parameters. Crucially, this empirical reward surface allows us to combine the online structural grounding of GRPO with the targeted offline refinement of DPO, outperforming either of the two algorithms in isolation, as well as Supervised Fine-Tuning (SFT) and In-Context Learning (ICL) baselines.

Our contributions are summarized as follows:
\begin{itemize}[leftmargin=*,noitemsep,topsep=0pt]
\item \textbf{Proxy-free online RL framework:} We introduce a paradigm for aligning LLMs that completely eliminates the instability and memory overhead of learned reward models, while preserving the ability for online exploration.
\item \textbf{Hybrid alignment strategy:} Our approach seamlessly bridges online structural grounding (GRPO) and offline preference sharpening (DPO), overcoming the inherent limitations of each method when used in isolation.
\item \textbf{Improved distillation:} Through this GRPO+DPO pipeline, we successfully distill subjective population preferences into a 1.5B parameter model, achieving perceptual parity with GPT-4o mini in a blind A/B listening test.
\end{itemize}

\section{Related Work}
\label{sec:related_work}

Our work positions itself at the intersection of semantic audio control, preference alignment, and RL in continuous action spaces. Traditionally, mapping natural language to audio parameters has been treated as a supervised regression problem. Early systems relied on fixed vocabulary-to-parameter mappings \cite{pardo2012building}, while recent approaches employ LLMs to interpret unconstrained text and estimate equalization (EQ) and reverberation (reverb) parameters \cite{chu2025text2fx, doh2025can}. Crucially, these regression-based methods implicitly favor a single parameter realization at inference time. Prior work \cite{Stylianou_2026} challenged this assumption, proposing that audio descriptors like ``warm'' map to distributions rather than point estimates. However, that study relied on small-scale supervised learning ($N=11$ per prompt). Our work scales this distributional premise ($N \approx 3,000$ per prompt) and transitions from supervised matching to RL, enabling on-policy optimization that fully exploits densely annotated data.

The standard LLM RL-tuning pipeline \cite{ouyang2022training} typically relies on PPO \cite{schulman2017proximalpolicyoptimizationalgorithms} optimized against a learned reward model. However, PPO requires training a separate Value Network to estimate the expected return. In subjective multimodal domains like audio preference, fitting a Value Network introduces an additional proxy objective, compounding the risk of reward misspecification \cite{casper2023openproblemsfundamentallimitations}. To circumvent the instability of critics and reward models, recent work has shifted toward direct alignment methods. Approaches like DPO, Identity Preference Optimization (IPO) \cite{azar2023generaltheoreticalparadigmunderstand}, Kahneman-Tversky Optimization (KTO) \cite{ethayarajh2024ktomodelalignmentprospect} and other variants \cite{chowdhury2024provably,liu2023statistical,ji2024towards}, optimize the policy directly from preference data without an explicit reward function. However, these methods are blind to the probability mass assigned to completions outside the preference dataset \cite{xu2024dpo, ivison2024unpacking}. As a result, the model may increase the probability of malformed or constraint-violating outputs during training, even though such outputs are never explicitly preferred. As we demonstrate, this lack of active exploration makes them brittle when the model must actively learn to maintain strict formatting constraints. While extensions like C-DPO \cite{liu2024enhancingllmsafetyconstrained} attempt to mitigate this by integrating explicit penalty margins, the lack of active exploration remains a bottleneck in continuous control tasks where the model must inherently learn the structural manifold.

GRPO \cite{shao2024deepseekmath} offers a middle ground between the heavy machinery of PPO and the static nature of DPO. Originally designed for reasoning tasks, GRPO removes the value network and instead uses the average reward of a group of generated rollouts as the baseline. By generating its own data during training, GRPO provides the exploration necessary to learn format constraints. Other variants such as Dr.GRPO \cite{liu2025understandingr1zeroliketrainingcritical} and DAPO \cite{yu2025dapoopensourcellmreinforcement} have proposed advanced advantage estimation and regularization techniques to improve upon this foundation. However, in this study we deliberately evaluate the vanilla instantiations of both DPO and GRPO. By omitting explicitly constrained decoding, customized validity penalties, or specific algorithmic extensions, we isolate the fundamental structural trade-off: the formatting failures of pure offline contrastive learning versus on-policy exploration of online methods.

To mitigate the individual weaknesses of online and offline methods, recent works in reasoning and multimodal generation have begun exploring hybrid pipelines that sequence online exploration with offline refinement \cite{li2025veripo, wu2025arm, tong2025delving}. Building on this emerging paradigm, our work adapts this hybrid stabilization strategy to continuous parameter estimation. Crucially, we do so while completely eliminating the reliance on the learned verifiers or reward models used in these prior works.

\section{Preliminaries and reward surface construction}
\label{sec:preliminaries}

To effectively align an LLM with subjective auditory preferences, we must first define the control space and construct a reliable reward signal that captures the nuance of human perception without relying on unstable proxy networks.

\subsection{The Beosonic control space}
\begin{wrapfigure}{r}{0.65\textwidth} 
    \vspace{-20pt}
    \centering
    \includegraphics[width=\linewidth]{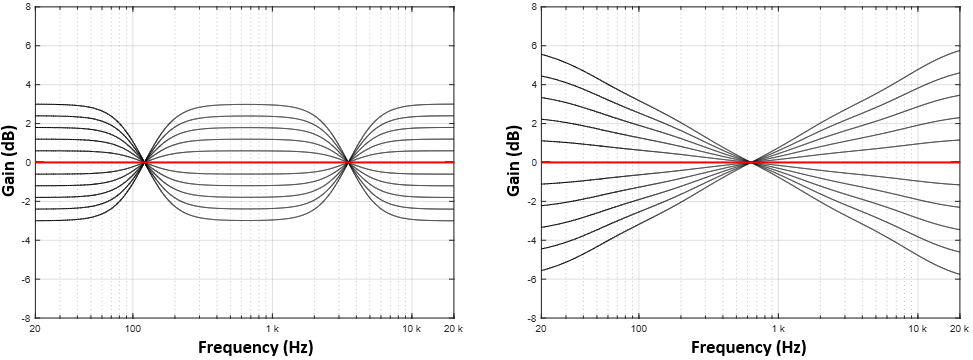}
    \caption{Filters of the EQ controller. Horizontal movement yields a ``smile curve'' (left); vertical applies a linear adjustment (right).}
    \label{fig:beosonic_filt}
\end{wrapfigure}

We ground our study in the ``Beosonic'' interface \cite{beosound_theatre_guide}, a continuous two-dimensional control plane $\mathcal{X} \in [-6, 6]^2$. This space reduces the EQ transfer function to 2 parameters, with each corresponding to the intensity of a distinct equalization filter ($\pm 6$ gain), as illustrated in Figure \ref{fig:beosonic_filt}. With this approach we effectively represent audio equalization through an interpretable 2D parameter space. Unlike discrete selection tasks, the model must predict a continuous coordinate $x \in [-6, 6]^2$.

\subsection{Dataset and motivation}
\label{subsec:dataset_motivation}
Our reward signal is the product of a large-scale comparison dataset, originally collected to evaluate the framework proposed in \cite{Stylianou_2026}. In this deployment, participants were presented with natural language prompts paired with audio variations generated by four distinct strategies: a deterministic Zero-Shot LLM (T2B), two distributional baselines (LoRA, RAG), and a Random agent. For each variation, users provided binary feedback indicating whether the audio satisfied the prompt.

\begin{wrapfigure}{r}{0.45\textwidth}
    \centering
    \vspace{-20pt}
    \includegraphics[width=\linewidth]{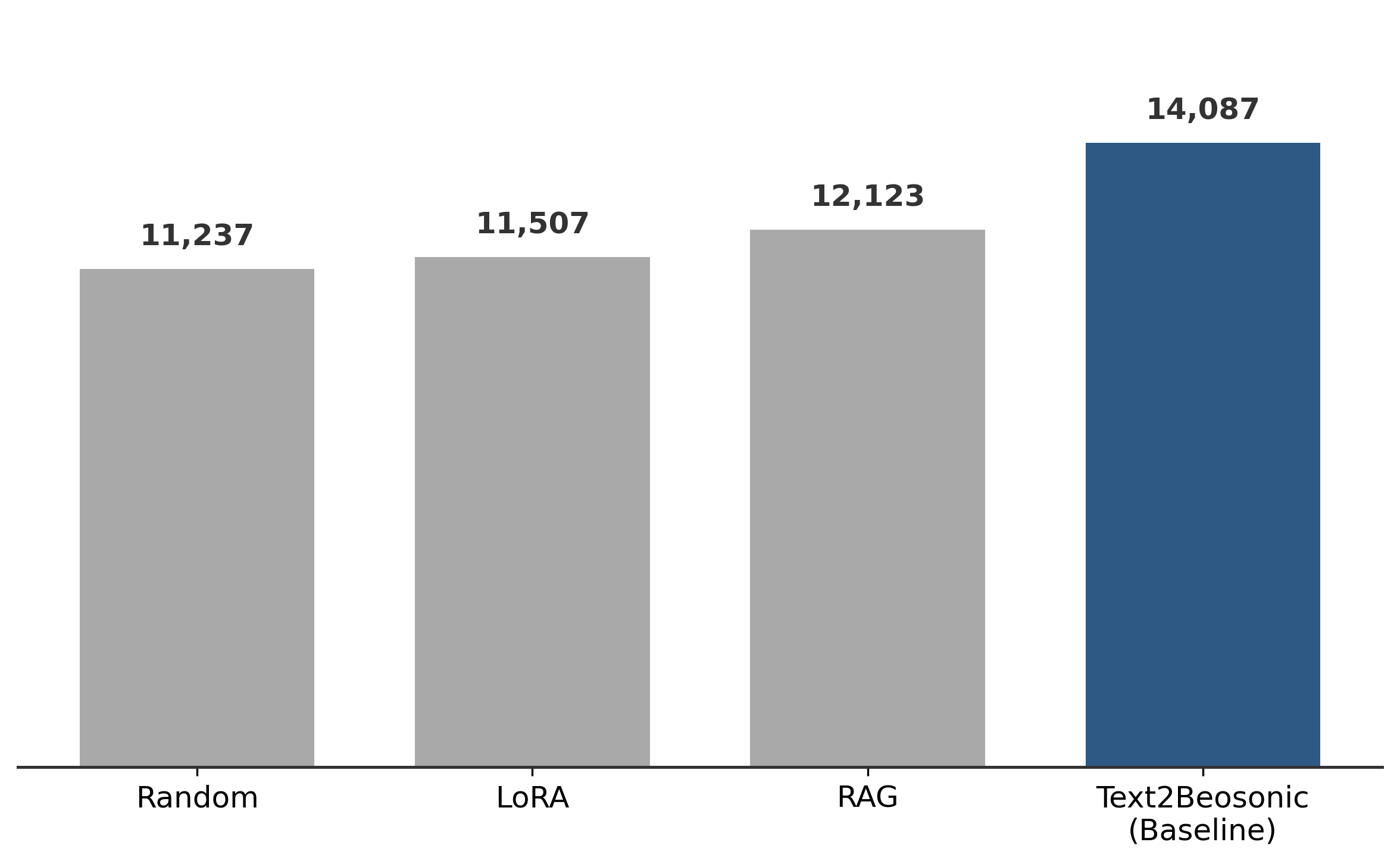}
    \caption{Total user selection counts by model type. The Zero-Shot baseline (T2B) outperforms other models (LoRA, RAG), indicating that previous attempts at distributional modeling failed to capture preference modes.}
    \label{fig:agent_counts}
    \vspace{-20pt}
\end{wrapfigure}

The resulting dataset comprises $N \approx 90,000$ interaction events (consisting of both selections and rejections). An analysis of the positive selections (Figure \ref{fig:agent_counts}) reveals a critical insight: The prompt-engineered Zero-Shot baseline (T2B) achieved the highest selection count ($14,087$), significantly outperforming the more complex distributional models. This outcome suggests that the prior distributional approaches, having been trained on extremely sparse data ($N=11$), failed to generalize effectively.

This observation motivates our RL approach: We require an objective that can explicitly optimize for the peaks of human satisfaction (Exploitation) while maintaining sufficient coverage of the space (Exploration).

\subsection{Probability of selection}
To construct the reward function $R(x|p)$ for a specific prompt $p$, we cannot simply estimate the density of high-rated points, as this would be biased by the sampling distribution of the generating models (e.g., if a model samples the center frequently, high density there might differ from high preference).

We therefore aim to estimate the Probability of Selection $P(\text{Selected}|x)$ for a given coordinate $x$. Let $\mathcal{D}_{all}$ be the set of all offered coordinates for prompt $p$, and $\mathcal{D}_{pref}\subseteq \mathcal{D}_{all}$ be the same set weighted by binary selection events. We estimate two densities using Reflective-KDE (defined in Subsection \ref{subsec:rKDE})

\begin{enumerate}[leftmargin=*,noitemsep,topsep=0pt]
    \item $\hat{f}_{total}(x)$: The density of the proposal distribution (where the points came from; $P(x)$).
    \item $\hat{f}_{pref}(x)$: The density of the users' preferred regions ($P(x|\text{Selected})$).
\end{enumerate}

By applying Bayes' theorem, the Probability of Selection is proportional to the ratio of the positive density to the total offered density:
\begin{equation}
\begin{aligned}
    S(x) &= P(\text{Selected} \mid x) = \frac{P(x \mid \text{Selected}) \, P(\text{Selected})}{P(x)} \propto \frac{\hat{f}_{\text{pref}}(x)}{\hat{f}_{\text{total}}(x) + \epsilon}
\end{aligned}
\end{equation}
where $\epsilon$ ensures numerical stability. This ratio effectively cancels out the sampling bias, isolating the underlying population preference.

\subsection{Reflective kernel density estimation}\label{subsec:rKDE}

Standard KDE suffers from boundary bias in bounded spaces, underestimating density near the edges (where user preferences often cluster, e.g., ``Max Bass'').
To correct this, we adopt Reflective-KDE \cite{schuster1985incorporating}, augmenting the dataset by mirroring each point $x_i$ across the boundaries of $\mathcal{X}$ into the 8 surrounding neighbor squares, creating a $3\times3$ grid of ``ghost'' points. The density estimate becomes:
\begin{equation}
    \hat{f}(x) = \frac{1}{9 n h^2} \sum_{i=1}^{n} \sum_{k=0}^{8} K \left( \frac{x - T_k(x_i)}{h} \right)
\end{equation}
where $T_k$ represents the reflection mappings ($T_0$ is the Identity) and $K$ is a Gaussian kernel. We set the bandwidth $h = 0.25 \cdot h_{\text{Scott}}$. This scaling factor on Scott's Rule \cite{scott2010scott} accounts for two specific properties of our space: a 0.5 factor to adjust for the extra point mass introduced by the reflective boundary, and an additional 0.5 factor to counteract the known tendency of Scott's Rule to over-smooth non-convex spaces. An example of the effect of different bandwidths is provided in Appendix \ref{sec:appendix_kde}.

\subsection{Global normalization and augmentation}

\begin{wrapfigure}{r}{0.55\textwidth}
    \vspace{-20pt}
    \centering
    \includegraphics[width=\linewidth]{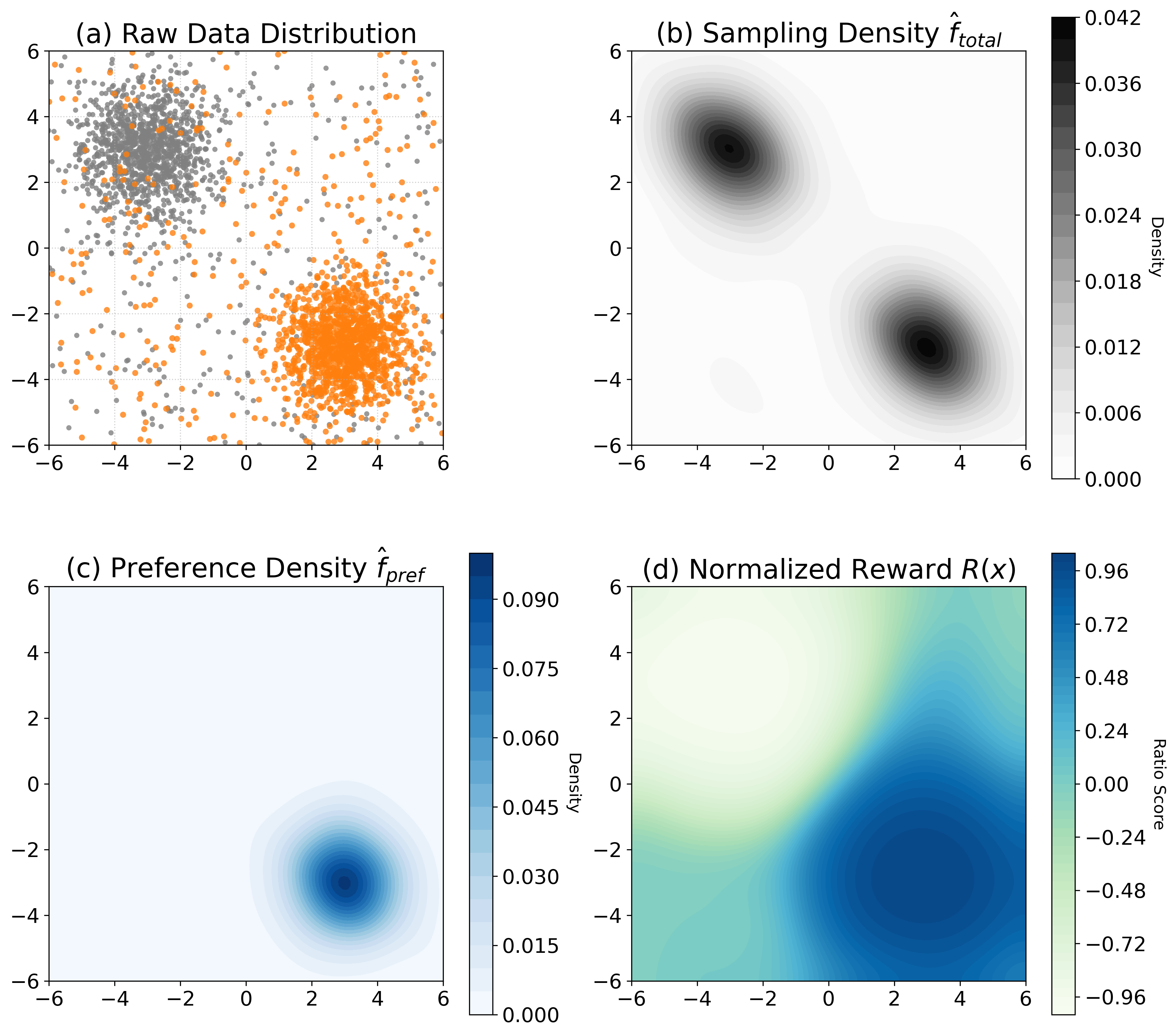}
    \caption{\textbf{Non-Parametric Reward.} (a) Sampling bias. (b) Total density $\hat{f}_{total}$. (c) Preferred density $\hat{f}_{pref}$. (d) Final normalized Reward Surface R(x).}
    \label{fig:norm_reward}
    \vspace{-40pt}
\end{wrapfigure}

To ensure the reward signal is consistent across different prompts, we perform a global normalization pass. We compute the global maximum ($S_{max}$) and minimum ($S_{min}$) density ratios across the entire dataset. The final reward $R(x)$ used for RL training is scaled to $[-1, 1]$:
\begin{equation}
    R(x) = 2 \cdot \frac{S(x) - S_{min}}{S_{max} - S_{min}} - 1
\end{equation}

The whole reward-surface construction process is exemplified in Figure \ref{fig:norm_reward}.

Finally, we employ semantic data augmentation. We assume that synonymous prompts (e.g., ``Make it warmer'' and ``Increase warmth'') share an identical preference manifold. We aggregate data for synonymous clusters and map them to the same reward surface.

\subsection{Evaluation splits}\label{subsec:eval_splits}
To assess generalization, we partition the prompts into three sets; a training set containing approximately $80\%$ of the prompts used for policy updates, a 10\% validation set consisting of the randomly held-out prompts drawn from the same distribution, and a manually isolated 10\%  Out-Of-Distribution (OOD) set containing a cluster of prompts related to ``Vocal Clarity.'' These OOD concepts are entirely unseen during training, allowing to evaluate the model's ability to extrapolate its learned preference geometry to novel semantic inputs.

\section{Alignment strategies in continuous spaces}
\label{sec:alignment}

Having constructed a non-parametric reward surface $R(x)$, our goal is to align an LLM policy $\pi_\theta$ to maximize the expected return $\mathbb{E}_{x \sim \pi_\theta}[R(x)]$. In this section, we analyze the structural limitations of standard offline alignment in continuous spaces and propose a hybrid pipeline that utilizes an online-aligned model to synthesize its own high-precision preference dataset.

\subsection{The challenge of continuous formatting}
Unlike standard chat models, our agent must output valid coordinates in a strict format (e.g., \texttt{[1.5, -3.2]}). A critical failure mode in this domain is \textit{format collapse}, where the model drifts into generating conversational text or invalid syntax.

We hypothesize that offline methods (i.e., DPO) are particularly susceptible to this. Since DPO is limited to the specific negative examples in the static dataset, it cannot penalize emergent structural failures. As the policy shifts, it may drift into generating novel malformed syntaxes that were not present in the pre-collected data, effectively bypassing the optimization constraints. Conversely, online methods (like GRPO) generate their own predictions during training. This allows the environment to provide immediate negative feedback ($R=-1.0$) for syntax errors, effectively ``grounding'' the model in the constraints of the control interface.

\subsection{Baseline 1: direct preference optimization}\label{subsec:DPO}
DPO trains the policy by increasing the margin between preferred responses and dispreferred ones:
\begin{equation}\label{eq:DPO}
    \mathcal{L}_{\text{DPO}} = -\mathbb{E}_{(p, x_w, x_l) \sim \mathcal{D}} \left[ \log \sigma \left( \hat{r}_\theta(p, x_w) - \hat{r}_\theta(p, x_l) \right) \right]
\end{equation}

where the implicit reward is defined as: $\hat{r}_\theta(p, x) = \beta \log\frac{\pi_\theta(x|p)}{\pi_{\text{ref}}(x|p)}.$

While efficient, DPO relies heavily on the Kullback-Leibler (KL)-divergence constraint (implicitly controlled by $\beta$) to prevent the model from deviating from the reference policy $\pi_{\text{ref}}$. In our domain, we construct the dataset $\mathcal{D}$ using the reward map $R(x)$, where $x_w$ and $x_l$ are pairs of coordinates such that $R(x_w|p) > R(x_l|p)$. More details are provided in Subsection \ref{subsec:hybrid}.

\subsection{Baseline 2: group relative policy optimization}
We adopt GRPO as our online framework. Unlike PPO, GRPO eliminates the value model. For each prompt $p$, the policy generates a group of $G$ outputs $\{x_1, \dots, x_G\}$. By setting $\rho_i(\theta)=\frac{\pi_\theta(x_i|p)}{\pi_{\text{old}}(x_i|p)}$, the objective maximizes:
\begin{small}
\begin{equation}
\begin{split}
    \mathcal{J}_{\text{GRPO}}(\theta) = \mathbb{E}_{\substack{p \sim P(\texttt{P}),  \{x_i\}_{i=1}^G \sim \pi_{\theta_{\text{old}}}}} \Bigg[ \frac{1}{G} \sum_{i=1}^G \bigg( 
    \underbrace{\min \Big( \rho_i(\theta) A_i, \text{clip}(\rho_i(\theta), 1-\epsilon, 1+\epsilon) A_i \Big)}_{\text{Clipped Surrogate Objective}}
    - \underbrace{\beta \mathbb{D}_{\text{KL}}(\pi_\theta || \pi_{\text{ref}})}_{\text{KL Penalty}} & \bigg) \Bigg]
\end{split}
\end{equation}
\end{small}
The advantage ${A}_i$ is computed by normalizing the rewards within the group. Crucially, if the model generates invalid format, the reward function returns $R=-1.0$. Since this penalty is included in the group, the model learns to avoid invalid syntax to achieve a positive advantage relative to its peers.

\subsection{Proposed method: Hybrid GRPO+DPO pipeline}\label{subsec:hybrid}
We observe that while GRPO excels at structural stability (learning \textit{how} to format), it may fail to find all the local maxima of the preference density (learning exactly \textit{what} different people want to hear). To resolve this, we propose a two-stage pipeline that combines the stability of online RL with the precision of offline RL.

\textbf{Stage 1: Structural Alignment (GRPO).}
We first train $\pi_\theta$ using GRPO against the Density Map. This transforms the generalist LLM into a robust control agent ($\pi_{GRPO}$) that reliably outputs coordinates in the correct manifold.

\textbf{Stage 2: Synthetic Preference Mining \& Peak Injection.}
We construct a synthetic DPO dataset $\mathcal{D}_{syn}$ using $\pi_{GRPO}$ as the generator. To ensure the dataset steers the model toward the local optima, we employ a peak injection strategy:
\begin{enumerate}[leftmargin=*,noitemsep,topsep=0pt]
    \item \textbf{On-Policy Mining:} We sample completions from $\pi_{GRPO}$. Those with high rewards ($R(x) > 0.8$) are added to the positive pool; malformed or low-reward outputs ($R(x) < -0.2$) are added to the negative pool.
    \item \textbf{Peak Injection:} If $\pi_{GRPO}$ fails to discover the top-K local maxima of the Density Map $R(x)$ for a given prompt, we analytically find the peak coordinates via grid search on $R(x)$ and inject them into the positive pool as ``Golden Truths.''
\end{enumerate}

The process is exactly the same as in standard DPO (Subsection \ref{subsec:DPO}), with $\pi_{ref}$ as the generator instead of $\pi_{GRPO}$.

\textbf{Stage 3: Refinement (DPO).}
We tune the model using DPO on $\mathcal{D}_{syn}$, setting $\pi_{ref} = \pi_{GRPO}$. By initializing with the GRPO model and training on injected expert peaks with relatively high $\beta$ values, the hybrid model achieves both high format compliance and improved preference matching.

\section{Experiments and results}
\label{sec:experiments}

We evaluate our alignment strategies on the OOD test set described in Section \ref{subsec:eval_splits}. This set contains semantic concepts unseen during training, requiring the model to generalize the geometric understanding of the preference space.

\subsection{Experimental setup}
\textbf{Models:} We utilize the \textbf{Qwen2.5-Instruct} family (0.5B and 1.5B parameters) \cite{qwen2.5} as our base models due to their strong instruction-following capabilities and open weights (Apache License 2.0).

\textbf{Baselines:} We compare against ICL baselines using Qwen2.5 and GPT-4o mini. To establish a rigorous performance floor, we inject domain knowledge into these baselines via a sophisticated system prompt containing an explicit semantic-to-coordinate mapping dictionary (e.g., ``warm'' $\to [0, -6]$), derived from audio attribute clustering literature \cite{francombe2017automatic}. In contrast, our RL-tuned models are agnostic to these expert definitions; they rely on a minimal system prompt and must learn the semantic-to-acoustic relationships purely from the reward signal.

\textbf{Metrics:}

\emph{Greedy Mean Relative Reward (GMRR):} We evaluate the models' performance on their most probable (\textbf{greedy}) completion for each prompt. Additionally, while training utilizes globally normalized rewards, evaluation requires measuring optimality \textbf{relative} to the specific prompt. We therefore revert to local normalization: For each prompt, the reward is rescaled such that the maximum of the density map is $1.0$ and the minimum is $0.0$.
    \begin{equation}
        \text{GMRR}(x) = \frac{S(x) - S_{min, local}}{S_{max, local} - S_{min, local}}.
    \end{equation}
Under this metric, a score of $1.0$ indicates the model found the absolute peak of human preference for that specific query, while the absolute low or malformed outputs are assigned $0.0$. 
    
\emph{Format Success Rate (Parse Rate):} The percentage of generated outputs that \textbf{strictly} adhere to the coordinate format \texttt{[x, y]}.

\subsection{The stability gap: Online vs. offline}
\label{subsec:stability_exp}

Our first experiment investigates the structural stability of the alignment algorithms. We sweep the KL-divergence penalty coefficient $\beta \in \{0.0, 0.1, 0.5, 1.0\}$ for both DPO (Offline) and GRPO (Online) and measure the Format Success Rate on the OOD test set.

As shown in Table \ref{tab:beta_analysis}, a critical failure mode emerges for DPO. A low $\beta$ value allows the model to deviate more aggressively from its initial state. During training, the objective of the algorithm is to maximize the margin between the positive and negative examples in the pre-developed preference dataset. This approach indeed increases the relative probability of generating a preferred example, \emph{but only compared to generating a dis-preferred one}. It is likely, especially with low KL regularization, that the overall probability of generating a positive example decreases during training. However, the training objective is still met as long as the overall probability of the negative counterpart decreases even faster. 
    
A different way to view this is that offline DPO tends to ``tunnel-vision'' on the two examples within each preference dataset entry, while the probabilities of all other completions are updated without any supervision. In contrast, GRPO maintains a 100\% format success rate even at $\beta=0.0$. Because the online agent generates its own data during training, any format deviation results in an immediate penalty ($R=-1$), forcing the policy to remain within the valid syntax manifold. Interestingly, excessively large $\beta$ values can over-constrain the model, hindering its ability to follow strict structure. 

\begin{wraptable}{r}{0.5\textwidth}
  \centering
  \vspace{-43pt}
  \caption{
    GMRR ({\scriptsize $\pm$ STD}) on the OOD test set and format success rate (\%). The top section compares the GRPO and DPO methods for different $\beta$ values. The bottom section provides ICL reference scores.
  }
  \label{tab:beta_analysis}
  \setlength{\tabcolsep}{3pt} 
  \resizebox{\linewidth}{!}{
  \begin{tabular}{@{} l l c c c c @{}}
    \toprule
    & & \multicolumn{2}{c}{\textbf{12-GRPO}} & \multicolumn{2}{c}{\textbf{DPO}} \\
    \cmidrule(lr){3-4} \cmidrule(lr){5-6}
    \textbf{$\beta$} & \textbf{Model} & \textbf{GMRR} & \textbf{Rate} & \textbf{GMRR} & \textbf{Rate} \\
    \midrule

    \multirow{2}{*}{0.0} 
      & Qwen2.5-0.5B & 0.44 $\pm$ {\scriptsize 0.08} & 100\% & 0.00 & 0\% \\
      & Qwen2.5-1.5B & 0.52 $\pm$ {\scriptsize 0.10} & 100\% & 0.00 & 0\% \\
    \midrule

    \multirow{2}{*}{0.1} 
      & Qwen2.5-0.5B & 0.49 $\pm$ {\scriptsize 0.14} & 100\% & 0.31 $\pm$ {\scriptsize 0.27} & 47\% \\
      & Qwen2.5-1.5B & 0.53 $\pm$ {\scriptsize 0.09} & 100\% & 0.00 & 0\% \\
    \midrule

    \multirow{2}{*}{0.5} 
      & Qwen2.5-0.5B & \textbf{0.52 $\pm$ {\scriptsize 0.10}} & 100\% & 0.33 $\pm$ {\scriptsize 0.22} & 81\% \\
      & Qwen2.5-1.5B & \textbf{0.53 $\pm$ {\scriptsize 0.09}} & 100\% & 0.32 $\pm$ {\scriptsize 0.27} & 69\% \\
    \midrule

    \multirow{2}{*}{1.0} 
      & Qwen2.5-0.5B & 0.50 $\pm$ {\scriptsize 0.10} & 94\% & 0.30 $\pm$ {\scriptsize 0.25} & 98\% \\
      & Qwen2.5-1.5B & 0.51 $\pm$ {\scriptsize 0.09} & 100\% & 0.07 $\pm$ {\scriptsize 0.23} & 10\% \\
    \midrule

    \multicolumn{2}{@{}l}{\textbf{ICL Reference Models}} & \multicolumn{2}{c}{\textbf{GMRR}} & \multicolumn{2}{c}{\textbf{Rate}} \\
    \cmidrule(l){1-2} \cmidrule(lr){3-4} \cmidrule(lr){5-6}
    
    \multirow{3}{*}{}
      & Qwen2.5-0.5B & \multicolumn{2}{c}{0.25 $\pm$ {\scriptsize 0.24}} & \multicolumn{2}{c}{54\%} \\
      & Qwen2.5-1.5B & \multicolumn{2}{c}{0.49 $\pm$ {\scriptsize 0.08}} & \multicolumn{2}{c}{100\%} \\
      & GPT-4o mini  & \multicolumn{2}{c}{\textbf{0.56} $\pm$ {\scriptsize 0.07}} & \multicolumn{2}{c}{100\%} \\
    \bottomrule
  \end{tabular}
  }
  \vspace{-30pt}
\end{wraptable}

\subsection{Scaling exploration}
We further analyze the impact of the group size $G$ (number of rollouts per prompt) on GRPO performance (Table \ref{tab:rollouts}). Increasing $G$ consistently improves the GMRR, saturating between $G=8$ and $G=16$. However, this improvement is relatively small, with even the 2-Rollout models outperforming their sophisticated ICL counterparts. This observation aligns with recent studies \cite{wu2025takestwogrposecretly}, which question the benefits of larger samples sizes relative to the computational burden.

\begin{table}[t]
  \centering
  
  \caption{
    Performance comparison of Qwen2.5 models under the GRPO algorithm with varying numbers of rollouts (sample size) and fixed $\beta$ value (0.5). Metrics shown are Greedy Mean Relative Reward ({\scriptsize $\pm$ Standard Deviation}) and format success rate (\%).
  }
  \label{tab:rollouts}
  \resizebox{\textwidth}{!}{%
  \begin{tabular}{@{} l *{5}{c c} @{}}
    \toprule
    & \multicolumn{2}{c}{\textbf{2 Rollouts}} & \multicolumn{2}{c}{\textbf{4 Rollouts}} & \multicolumn{2}{c}{\textbf{8 Rollouts}} & \multicolumn{2}{c}{\textbf{16 Rollouts}} & \multicolumn{2}{c}{\textbf{32 Rollouts}} \\
    \cmidrule(lr){2-3} \cmidrule(lr){4-5} \cmidrule(lr){6-7} \cmidrule(lr){8-9} \cmidrule(lr){10-11}
    
    \textbf{Model} & {\textbf{GMRR}} & {\textbf{Parse Rate}} & {\textbf{GMRR}} & {\textbf{Parse Rate}} & {\textbf{GMRR}} & {\textbf{Parse Rate}} & {\textbf{GMRR}} & {\textbf{Parse Rate}} & {\textbf{GMRR}} & {\textbf{Parse Rate}} \\
    \midrule

    Qwen2.5-0.5B & 0.47 $\pm$ \scriptsize{0.17} & 81\% & 0.50 $\pm$ \scriptsize{0.14} & 93\% & \textbf{0.53 $\pm$ \scriptsize{0.10}} & 100\% & 0.51 $\pm$ \scriptsize{0.10} & 100\% & 0.48 $\pm$ \scriptsize{0.14} & 96\% \\
    Qwen2.5-1.5B & 0.50 $\pm$ \scriptsize{0.08} & 100\% & 0.51 $\pm$ \scriptsize{0.09} & 100\% & 0.51 $\pm$ \scriptsize{0.09} & 100\% & \textbf{0.52 $\pm$ \scriptsize{0.09}} & 100\% & 0.52 $\pm$ \scriptsize{0.09} & 100\% \\
    \bottomrule
  \end{tabular}
  }
\end{table}

\subsection{Hybrid pipeline performance}
\label{subsec:Hybrid_performance}
Finally, we evaluate the proposed \textbf{hybrid GRPO+DPO} pipeline. We select the best GRPO configuration ($\beta=0.5$) as the Reference and use DPO ($\beta = 1.0$) on the synthetically mined ``Peak Injection'' dataset (Subsection \ref{subsec:hybrid}) across different rollouts.

Results in Table \ref{tab:hybrid_results} demonstrate that for the 0.5B model, GMRR jumps from $0.53$ (Base GRPO) to $\textbf{0.56}$ (hybrid), matching the GPT-4o baseline at the cost of some inconsistency (97\% Parse Rate). The 1.5B model achieves a GMRR of $\textbf{0.60}$ without compromising format compliance. However, the hybrid approach exhibits performance fluctuations across rollout configurations. To confirm this reflects hyperparameter sensitivity rather than algorithmic instability, we evaluated the optimal configuration across four independent seeds, achieving a consistent mean GMRR of 0.58
(see Appendix \ref{sec:appendix_multiseed}).

\begin{table*}[ t]
\centering
\caption{
        Performance comparison of Qwen2.5 models using GRPO (Baseline) versus the combined GRPO + DPO algorithm. The $\beta$ values for GRPO and DPO are fixed at 0.5 and 1.0 respectively.
    }
\label{tab:hybrid_results}
\resizebox{\textwidth}{!}{%
\begin{tabular}{@{} l c l c c l @{}}
\toprule
\textbf{Model} & \textbf{Rollouts} & \textbf{Approach} & \textbf{GMRR} & \textbf{Parse Rate (\%)} & \textbf{Observation} \\ 
\midrule

\rowcolor{mygray} \cellcolor{white} & \multirow{2}{*}{2} & GRPO & \textbf{0.47} $\pm$ \scriptsize{0.17} & 81 &  \\
 & & \textbf{GRPO+DPO} & 0.44 $\pm$ \scriptsize{0.32} & \textbf{88} & Drop in GMRR, improvement in parsing rate \\ 
\cmidrule(lr){2-6}

\rowcolor{mygray} \cellcolor{white} & \multirow{2}{*}{4} & GRPO & 0.50 $\pm$ \scriptsize{0.14} & \textbf{93} &  \\
 & & \textbf{GRPO+DPO} & \textbf{0.56} $\pm$ \scriptsize{0.26} & 73 & Preference sharpened, but stability drops \\ 
\cmidrule(lr){2-6}

\rowcolor{mygray} \cellcolor{white} & \multirow{2}{*}{8} & GRPO & \textbf{0.53} $\pm$ \scriptsize{0.10} & 100 &  \\
 & & \textbf{GRPO+DPO} & 0.48 $\pm$ \scriptsize{0.25} & 100 & Drop in GMRR \\ 
\cmidrule(lr){2-6}

\rowcolor{mygray} \cellcolor{white} & \multirow{2}{*}{16} & GRPO & \textbf{0.51} $\pm$ \scriptsize{0.10} & \textbf{100} &  \\
 & & \textbf{GRPO+DPO} & 0.49 $\pm$ \scriptsize{0.32} & 74 & Drop in both metrics \\ 
\cmidrule(lr){2-6}

\rowcolor{mygray} \cellcolor{white} & \multirow{2}{*}{32} & GRPO & 0.48 $\pm$ \scriptsize{0.14} & 96 &  \\
\multirow{-10}{*}{\textbf{Qwen2.5-0.5B}} & & \textbf{GRPO+DPO} & \textbf{0.56} $\pm$ \scriptsize{0.18} & \textbf{97} & Improvement in both metrics \\ 
\midrule


\rowcolor{mygray} \cellcolor{white} & \multirow{2}{*}{2} & GRPO & \textbf{0.50} $\pm$ \scriptsize{0.08} & 100 &  \\
 & & \textbf{GRPO+DPO} & 0.27 $\pm$ \scriptsize{0.24} & 100 & GMRR Drop \\ 
\cmidrule(lr){2-6}

\rowcolor{mygray} \cellcolor{white} & \multirow{2}{*}{4} & GRPO & 0.51 $\pm$ \scriptsize{0.09} & \textbf{100} &  \\
 & & \textbf{GRPO+DPO} & \textbf{0.56} $\pm$ \scriptsize{0.18} & 78 & Improvement but stability degradation hurts GMRR \\ 
\cmidrule(lr){2-6}

\rowcolor{mygray} \cellcolor{white} & \multirow{2}{*}{8} & GRPO & \textbf{0.51} $\pm$ \scriptsize{0.09} & 100 &  \\
 & & \textbf{GRPO+DPO} & 0.40 $\pm$ \scriptsize{0.17} & 100 & GMRR Drop \\ 
\cmidrule(lr){2-6}

\rowcolor{mygray} \cellcolor{white} & \multirow{2}{*}{16} & GRPO & 0.52 $\pm$ \scriptsize{0.09} & 100 &  \\
 & & \textbf{GRPO+DPO} & \textbf{0.60} $\pm$ \scriptsize{0.17} & 100 & \textbf{Strongest Improvement (GMRR \& Rate)} \\ 
\cmidrule(lr){2-6}

\rowcolor{mygray} \cellcolor{white} & \multirow{2}{*}{32} & GRPO & \textbf{0.52} $\pm$ \scriptsize{0.09} & 100 &  \\
\multirow{-10}{*}{\textbf{Qwen2.5-1.5B}} & & \textbf{GRPO+DPO} & 0.48 $\pm$ \scriptsize{0.13} & 100 & GMRR Drop \\ 
\bottomrule
\end{tabular}
}
\end{table*}

To assess if the preference manifold could be learned via supervision, we also trained an SFT baseline with Low Rank Adaptation \cite{hu2021loralowrankadaptationlarge} (LoRA) on the ``peaks'' of the training prompts. This model achieved a GMRR of 0.45, performing worse than the base model with In-Context Learning (0.49). This suggests that SFT leads to overfitting, degrading the model's ability to extrapolate to OOD semantic concepts. Table \ref{tab:summary_comparison} provides a performance summary of the different methods for Qwen2.5-1.5B.

\section{Subjective evaluation}
\label{sec:subjective}

While our objective metrics demonstrate the convergence of the proposed RL approach, the ultimate validation of an audio system lies in listener perception. We conducted a blind A/B listening test to compare our best-performing hybrid model (Qwen2.5-1.5B) against the baseline (GPT-4o mini via In-Context Learning). Both models controlled the exact same equalization system, ensuring that any perceived differences result purely from the parameters selected by the LLMs.

\begin{wraptable}{r}{0.45\textwidth}
    \vspace{-35pt}
    \centering
    \caption{Performance summary of Qwen2.5-1.5B on the OOD Test Set. \textbf{GRPO} improves generalization compared to \textbf{SFT} via online exploration, while the \textbf{hybrid GRPO+DPO} pipeline achieves the highest performance.}
    \label{tab:summary_comparison}
    \setlength{\tabcolsep}{3pt} 
    \small
    \begin{tabular}{l l c c}
    \toprule
    \textbf{Model} & \textbf{Approach} & \textbf{GMRR} & \textbf{Rate} \\
    \midrule
    1.5B & SFT (Peak) & 0.45 $\pm$ {\scriptsize 0.12} & 100\% \\
    1.5B & ICL & 0.49 $\pm$ {\scriptsize 0.08} & 100\% \\
    1.5B & GRPO & 0.53 $\pm$ {\scriptsize 0.09} & 100\% \\
    1.5B & Hybrid & \textbf{0.60} $\pm$ \scriptsize{0.17} & 100\% \\
    \bottomrule
    \end{tabular}
    \vspace{-10pt}
\end{wraptable}

\subsection{Unconstrained data collection and protocol}
To ensure the evaluation reflected real-world interaction, we utilized a two-phase study with human listeners. First, we recruited 12 participants for a prompt-collection phase. We employed an unconstrained elicitation protocol: Participants listened to audio clips and were asked to request \textit{any} changes they desired without being informed of the system's technical limitations (i.e., that it was an equalizer). This design choice was deliberate to avoid \textit{priming bias}, ensuring that users did not subconsciously tailor their requests to perceived system capabilities (e.g., only asking for ``more bass''). As expected, this yielded highly Out-Of-Distribution requests (e.g., volume control, removing instruments, or changing speaker identity). After filtering out these functional system requests, a set of 30 prompts was retained. 

In the second phase, 11 of the original participants returned to evaluate the models' responses on these 30 prompts. For each prompt, participants listened to audio processed by both the hybrid RL model and the GPT-4o mini baseline in a randomized, blinded interface. They rated how well the resulting audio satisfied the original request on a 5-point Likert scale (1=``Very Poorly'' to 5=``Very Well'').

\begin{wraptable}{r}{0.43\textwidth}
    \vspace{-40pt}
    \centering
    \caption{Subjective evaluation on Likert score ({\scriptsize $\pm$ Standard Error}). The $p$-value is calculated via the Linear Mixed-Effects Model.}
    \label{tab:subjective_results}
    \setlength{\tabcolsep}{3pt} 
    \small
    \begin{tabular}{l c c c}
        \toprule
        \textbf{Method} & \textbf{Score (1--5)} & \textbf{Win/Tie \%} & \textbf{\textit{p}-value} \\
        \midrule
        GPT-4o & 2.92 \scriptsize{$\pm0.06$} & 72.1\% & \multirow{2}{*}{0.095} \\
        \textbf{Hybrid} & \textbf{3.04} \scriptsize{$\pm0.06$} & \textbf{77.9\%} & \\
        \bottomrule
    \end{tabular}
    \vspace{-20pt}
\end{wraptable}

\subsection{Quantitative results}
The aggregate results (Table \ref{tab:subjective_results}) indicate that our specialized RL model performs at parity with the previously-best generalist baseline. More specifically:

\textbf{Mean Preference Score:} The 1.5B RL model achieved a mean rating of \textbf{3.04}, outperforming the GPT-4o mini baseline (2.92). This suggests that the RL model handles ambiguous, OOD prompts gracefully.

\textbf{Win/Tie Rate:} In head-to-head comparisons, the RL model was preferred or rated as equivalent to the baseline in \textbf{77.9\%} of the 330 trials (11 participants $\times$ 30 prompts). 

\textbf{Statistical Analysis:} Because a total of  660  evaluations (two per trial) were presented individually in randomized order, we analyzed the results using a Linear Mixed-Effects Model (LMM) with crossed random effects for both participant and prompt to account for repeated measures \cite{baayen2008mixed}. The LMM revealed a non-significant main effect between the models ($p = 0.095$). To formally establish perceptual parity between the 1.5B model and the baseline, we subsequently conducted an equivalence test using the Two One-Sided Tests (TOST) procedure \cite{lakens2017equivalence}. Detailed methodology, the full LMM summary matrix, and the TOST confidence intervals demonstrating statistical parity are provided in Appendix \ref{sec:app_stats}.

The critical implication of these results is efficiency: The proposed Density-Based RL pipeline successfully distilled the necessary audio control capabilities into a model that is not utilizing expert knowledge and is orders of magnitude smaller than its competitor, achieving equivalent user satisfaction without the need for large inference compute or proprietary APIs.

\section{Discussion and limitations}
\label{sec:discussion}

Our results suggest a paradigm shift for LLM alignment: Rather than training unstable proxy reward models, we can leverage the ``wisdom of the crowd'' directly via non-parametric density maps. However, this approach entails specific constraints that merit discussion.

\paragraph{The curse of dimensionality}
Our Reflective-KDE approach is highly effective for low-dimensional spaces (e.g., the 2D Beosonic interface). However, KDE suffers from the curse of dimensionality \cite{bellman1957dynamic}. As the control space expands (e.g., additional acoustic parameters or a 6-DoF robotic arm), the data required to populate a dense reward map grows exponentially. Scaling this framework to high-dimensional action spaces will likely require latent variable models (e.g., Variational Auto Encoders \cite{kingma2013auto}) to compress the control space into a tractable lower-dimensional manifold before applying our density-based RL.

\paragraph{Text-only conditioning}
The current system aligns audio parameters solely based on the semantic intent of the text prompt (e.g., ``Make it warm''). It does not analyze the input audio signal ($P(\text{Selected}|x, \text{Audio}_{ID})$). While this enables a lightweight, sparse (responds when requested), privacy-preserving inference (no audio upload required), it limits the system's context awareness. A ``warm'' adjustment for a podcast could differ significantly from a ``warm'' adjustment for techno music. Future iterations could integrate audio embeddings (e.g., CLAP \cite{elizalde2023clap} features) into the policy to enable content-adaptive equalization.

\paragraph{Consensus vs. personalization}
Finally, while our density map method effectively models the population's preference space, it does not yet solve the problem of individual personalization. By optimizing for the peaks of the density map, the hybrid model acts as a ``Mode-Seeking'' agent--it generates the setting that could satisfy \textit{any} user. It does not yet have the capacity to adapt to user-specific hearing profiles. Extending this framework to conditioned density estimation ($P(\text{Selected}|x, \text{User}_{ID})$) remains a promising and natural avenue for future research.

\section{Conclusion}
\label{sec:conclusion}

In this work, we presented a proxy-free reinforcement learning framework for continuous parameter estimation. By applying Reflective Kernel Density Estimation to large-scale user data ($N \approx 90,000$), we constructed non-parametric preference density maps that serve as a stable, empirical reward surface. This approach completely eliminates the memory overhead and reward-hacking vulnerabilities of learned proxy reward models, while preserving the ability to perform true online exploration. 

Crucially, this continuous reward surface provides the flexibility to employ and combine distinct alignment methodologies. It seamlessly enables both the active, on-policy exploration of GRPO and the targeted, contrastive optimization of DPO. By leveraging this capability to build a hybrid pipeline, we successfully aligned a highly efficient 1.5B-parameter model, achieving perceptual parity with a heavily prompt-engineered GPT-4o mini baseline in a blind A/B listening test.

\bibliographystyle{plainnat}
\bibliography{bibliography}

@article{doh2025can,
  title={Can Large Language Models Predict Audio Effects Parameters from Natural Language?},
  author={Doh, Seungheon and Koo, Junghyun and Mart{\'\i}nez-Ram{\'\i}rez, Marco A and Liao, Wei-Hsiang and Nam, Juhan and Mitsufuji, Yuki},
  journal={arXiv preprint arXiv:2505.20770},
  year={2025}
}

@inproceedings{chu2025text2fx,
  title={Text2fx: Harnessing CLAP embeddings for text-guided audio effects},
  author={Chu, Annie and O’Reilly, Patrick and Barnett, Julia and Pardo, Bryan},
  booktitle={ICASSP 2025-2025 IEEE International Conference on Acoustics, Speech and Signal Processing (ICASSP)},
  pages={1--5},
  year={2025},
  organization={IEEE}
}

@article{rafailov2023direct,
  title={Direct preference optimization: Your language model is secretly a reward model},
  author={Rafailov, Rafael and Sharma, Archit and Mitchell, Eric and Manning, Christopher D and Ermon, Stefano and Finn, Chelsea},
  journal={Advances in neural information processing systems},
  volume={36},
  pages={53728--53741},
  year={2023}
}

@article{shao2024deepseekmath,
  title={Deepseekmath: Pushing the limits of mathematical reasoning in open language models},
  author={Shao, Zhihong and Wang, Peiyi and Zhu, Qihao and Xu, Runxin and Song, Junxiao and Bi, Xiao and Zhang, Haowei and Zhang, Mingchuan and Li, YK and Wu, Yang and others},
  journal={arXiv preprint arXiv:2402.03300},
  year={2024}
}

@article{ouyang2022training,
  title={Training language models to follow instructions with human feedback},
  author={Ouyang, Long and Wu, Jeffrey and Jiang, Xu and Almeida, Diogo and Wainwright, Carroll and Mishkin, Pamela and Zhang, Chong and Agarwal, Sandhini and Slama, Katarina and Ray, Alex and others},
  journal={Advances in neural information processing systems},
  volume={35},
  pages={27730--27744},
  year={2022}
}

@inproceedings{gao2023scaling,
  title={Scaling laws for reward model overoptimization},
  author={Gao, Leo and Schulman, John and Hilton, Jacob},
  booktitle={International Conference on Machine Learning},
  pages={10835--10866},
  year={2023},
  organization={PMLR}
}

@article{schuster1985incorporating,
  title={Incorporating support constraints into nonparametric estimators of densities},
  author={Schuster, Eugene F},
  journal={Communications in Statistics-Theory and methods},
  volume={14},
  number={5},
  pages={1123--1136},
  year={1985},
  publisher={Taylor \& Francis}
}

@misc{beosound_theatre_guide,
  author       = {Bang\&Olufsen},
  title        = {Beosound Beovision Theatre: Technical Sound Guide},
  year         = {2023},
 

}

@article{Stylianou_2026,
   title={One Prompt, Many Sounds: Modeling Listener Variability in LLM-Based Equalization},
   ISSN={1941-0484},
   url={http://dx.doi.org/10.1109/JSTSP.2026.3710145},
   DOI={10.1109/jstsp.2026.3710145},
   journal={IEEE Journal of Selected Topics in Signal Processing},
   publisher={Institute of Electrical and Electronics Engineers (IEEE)},
   author={Stylianou, Ioannis and Francombe, Jon and Mart´ ınez-Nuevo, Pablo and Shepstone, Sven Ewan and Tan, Zheng-Hua},
   year={2026},
   pages={1–13} }

@misc{schulman2017proximalpolicyoptimizationalgorithms,
      title={Proximal Policy Optimization Algorithms}, 
      author={John Schulman and Filip Wolski and Prafulla Dhariwal and Alec Radford and Oleg Klimov},
      year={2017},
      eprint={1707.06347},
      archivePrefix={arXiv},
      primaryClass={cs.LG},
      url={https://arxiv.org/abs/1707.06347}, 
}

@misc{qwen2.5,
    title = {Qwen2.5: A Party of Foundation Models},
    url = {https://qwenlm.github.io/blog/qwen2.5/},
    author = {Qwen Team},
    month = {September},
    year = {2024}
}

@inproceedings{francombe2017automatic,
  title={Automatic text clustering for audio attribute elicitation experiment responses},
  author={Francombe, Jon and Brookes, Timothy and Mason, Russell},
  booktitle={AES 143rd Convention},
  year={2017},
  organization={Audio Engineering Society}
}

@inproceedings{elizalde2023clap,
  title={CLAP learning audio concepts from natural language supervision},
  author={Elizalde, Benjamin and Deshmukh, Soham and Al Ismail, Mahmoud and Wang, Huaming},
  booktitle={ICASSP 2023-2023 IEEE International Conference on Acoustics, Speech and Signal Processing (ICASSP)},
  pages={1--5},
  year={2023},
  organization={IEEE}
}

@article{kingma2013auto,
  title={Auto-encoding variational bayes},
  author={Kingma, Diederik P and Welling, Max},
  journal={arXiv preprint arXiv:1312.6114},
  year={2013}
}

@misc{wu2025takestwogrposecretly,
      title={It Takes Two: Your GRPO Is Secretly DPO}, 
      author={Yihong Wu and Liheng Ma and Lei Ding and Muzhi Li and Xinyu Wang and Kejia Chen and Zhan Su and Zhanguang Zhang and Chenyang Huang and Yingxue Zhang and Mark Coates and Jian-Yun Nie},
      year={2025},
      eprint={2510.00977},
      archivePrefix={arXiv},
      primaryClass={cs.LG},
      url={https://arxiv.org/abs/2510.00977}, 
}

@inproceedings{pardo2012building,
  title={Building a personalized audio equalizer interface with transfer learning and active learning},
  author={Pardo, Bryan and Little, David and Gergle, Darren},
  booktitle={Proceedings of the second international ACM workshop on Music information retrieval with user-centered and multimodal strategies},
  pages={13--18},
  year={2012}
}

@misc{ethayarajh2024ktomodelalignmentprospect,
      title={KTO: Model Alignment as Prospect Theoretic Optimization}, 
      author={Kawin Ethayarajh and Winnie Xu and Niklas Muennighoff and Dan Jurafsky and Douwe Kiela},
      year={2024},
      eprint={2402.01306},
      archivePrefix={arXiv},
      primaryClass={cs.LG},
      url={https://arxiv.org/abs/2402.01306}, 
}

@misc{azar2023generaltheoreticalparadigmunderstand,
      title={A General Theoretical Paradigm to Understand Learning from Human Preferences}, 
      author={Mohammad Gheshlaghi Azar and Mark Rowland and Bilal Piot and Daniel Guo and Daniele Calandriello and Michal Valko and Rémi Munos},
      year={2023},
      eprint={2310.12036},
      archivePrefix={arXiv},
      primaryClass={cs.AI},
      url={https://arxiv.org/abs/2310.12036}, 
}

@misc{casper2023openproblemsfundamentallimitations,
      title={Open Problems and Fundamental Limitations of Reinforcement Learning from Human Feedback}, 
      author={Stephen Casper and Xander Davies and Claudia Shi and Thomas Krendl Gilbert and Jérémy Scheurer and Javier Rando and Rachel Freedman and Tomasz Korbak and David Lindner and Pedro Freire and Tony Wang and Samuel Marks and Charbel-Raphaël Segerie and Micah Carroll and Andi Peng and Phillip Christoffersen and Mehul Damani and Stewart Slocum and Usman Anwar and Anand Siththaranjan and Max Nadeau and Eric J. Michaud and Jacob Pfau and Dmitrii Krasheninnikov and Xin Chen and Lauro Langosco and Peter Hase and Erdem Bıyık and Anca Dragan and David Krueger and Dorsa Sadigh and Dylan Hadfield-Menell},
      year={2023},
      eprint={2307.15217},
      archivePrefix={arXiv},
      primaryClass={cs.AI},
      url={https://arxiv.org/abs/2307.15217}, 
}

@misc{hu2021loralowrankadaptationlarge,
      title={LoRA: Low-Rank Adaptation of Large Language Models}, 
      author={Edward J. Hu and Yelong Shen and Phillip Wallis and Zeyuan Allen-Zhu and Yuanzhi Li and Shean Wang and Lu Wang and Weizhu Chen},
      year={2021},
      eprint={2106.09685},
      archivePrefix={arXiv},
      primaryClass={cs.CL},
      url={https://arxiv.org/abs/2106.09685}, 
}

@article{bellman1957dynamic,
  title={Dynamic programming, princeton univ},
  author={Bellman, Richard},
  journal={Press Princeton, New Jersey},
  volume={39},
  year={1957}
}

@article{scott2010scott,
  title={Scott's rule},
  author={Scott, David W},
  journal={Wiley Interdisciplinary Reviews: Computational Statistics},
  volume={2},
  number={4},
  pages={497--502},
  year={2010},
  publisher={Wiley Online Library}
}

@misc{liu2024enhancingllmsafetyconstrained,
      title={Enhancing LLM Safety via Constrained Direct Preference Optimization}, 
      author={Zixuan Liu and Xiaolin Sun and Zizhan Zheng},
      year={2024},
      eprint={2403.02475},
      archivePrefix={arXiv},
      primaryClass={cs.LG},
      url={https://arxiv.org/abs/2403.02475}, 
}

@misc{liu2025understandingr1zeroliketrainingcritical,
      title={Understanding R1-Zero-Like Training: A Critical Perspective}, 
      author={Zichen Liu and Changyu Chen and Wenjun Li and Penghui Qi and Tianyu Pang and Chao Du and Wee Sun Lee and Min Lin},
      year={2025},
      eprint={2503.20783},
      archivePrefix={arXiv},
      primaryClass={cs.LG},
      url={https://arxiv.org/abs/2503.20783}, 
}

@misc{yu2025dapoopensourcellmreinforcement,
      title={DAPO: An Open-Source LLM Reinforcement Learning System at Scale}, 
      author={Qiying Yu and Zheng Zhang and Ruofei Zhu and Yufeng Yuan and Xiaochen Zuo and Yu Yue and Weinan Dai and Tiantian Fan and Gaohong Liu and Lingjun Liu and Xin Liu and Haibin Lin and Zhiqi Lin and Bole Ma and Guangming Sheng and Yuxuan Tong and Chi Zhang and Mofan Zhang and Wang Zhang and Hang Zhu and Jinhua Zhu and Jiaze Chen and Jiangjie Chen and Chengyi Wang and Hongli Yu and Yuxuan Song and Xiangpeng Wei and Hao Zhou and Jingjing Liu and Wei-Ying Ma and Ya-Qin Zhang and Lin Yan and Mu Qiao and Yonghui Wu and Mingxuan Wang},
      year={2025},
      eprint={2503.14476},
      archivePrefix={arXiv},
      primaryClass={cs.LG},
      url={https://arxiv.org/abs/2503.14476}, 
}

@article{frechet1906quelques,
  title={Sur quelques points du calcul fonctionnel},
  author={Fr{\'e}chet, M Maurice},
  journal={Rendiconti del Circolo Matematico di Palermo (1884-1940)},
  volume={22},
  number={1},
  pages={1--72},
  year={1906},
  publisher={Springer Milan Milan}
}

@article{chowdhury2024provably,
  title={Provably robust dpo: Aligning language models with noisy feedback},
  author={Chowdhury, Sayak Ray and Kini, Anush and Natarajan, Nagarajan},
  journal={arXiv preprint arXiv:2403.00409},
  year={2024}
}

@article{liu2023statistical,
  title={Statistical rejection sampling improves preference optimization},
  author={Liu, Tianqi and Zhao, Yao and Joshi, Rishabh and Khalman, Misha and Saleh, Mohammad and Liu, Peter J and Liu, Jialu},
  journal={arXiv preprint arXiv:2309.06657},
  year={2023}
}

@article{ji2024towards,
  title={Towards efficient exact optimization of language model alignment},
  author={Ji, Haozhe and Lu, Cheng and Niu, Yilin and Ke, Pei and Wang, Hongning and Zhu, Jun and Tang, Jie and Huang, Minlie},
  journal={arXiv preprint arXiv:2402.00856},
  year={2024}
}

@article{lakens2017equivalence,
  title={Equivalence tests: A practical primer for t tests, correlations, and meta-analyses},
  author={Lakens, Dani{\"e}l},
  journal={Social psychological and personality science},
  volume={8},
  number={4},
  pages={355--362},
  year={2017},
  publisher={Sage Publications Sage CA: Los Angeles, CA}
}

@article{seabold2010statsmodels,
  title={Statsmodels: econometric and statistical modeling with python.},
  author={Seabold, Skipper and Perktold, Josef and others},
  journal={scipy},
  volume={7},
  number={1},
  pages={92--96},
  year={2010}
}

@article{baayen2008mixed,
  title={Mixed-effects modeling with crossed random effects for subjects and items},
  author={Baayen, R Harald and Davidson, Douglas J and Bates, Douglas M},
  journal={Journal of memory and language},
  volume={59},
  number={4},
  pages={390--412},
  year={2008},
  publisher={Elsevier}
}

@article{li2025veripo,
  title={Veripo: Cultivating long reasoning in video-llms via verifier-gudied iterative policy optimization},
  author={Li, Yunxin and Chen, Xinyu and Li, Zitao and Liu, Zhenyu and Wang, Longyue and Luo, Wenhan and Hu, Baotian and Zhang, Min},
  journal={arXiv preprint arXiv:2505.19000},
  year={2025}
}

@article{wu2025arm,
  title={Arm: Adaptive reasoning model},
  author={Wu, Siye and Xie, Jian and Zhang, Yikai and Chen, Aili and Zhang, Kai and Su, Yu and Xiao, Yanghua},
  journal={arXiv preprint arXiv:2505.20258},
  year={2025}
}

@article{tong2025delving,
  title={Delving into rl for image generation with cot: A study on dpo vs. grpo},
  author={Tong, Chengzhuo and Guo, Ziyu and Zhang, Renrui and Shan, Wenyu and Wei, Xinyu and Xing, Zhenghao and Li, Hongsheng and Heng, Pheng-Ann},
  journal={arXiv preprint arXiv:2505.17017},
  year={2025}
}

@article{xu2024dpo,
  title={Is dpo superior to ppo for llm alignment? a comprehensive study},
  author={Xu, Shusheng and Fu, Wei and Gao, Jiaxuan and Ye, Wenjie and Liu, Weilin and Mei, Zhiyu and Wang, Guangju and Yu, Chao and Wu, Yi},
  journal={arXiv preprint arXiv:2404.10719},
  year={2024}
}

@article{ivison2024unpacking,
  title={Unpacking dpo and ppo: Disentangling best practices for learning from preference feedback},
  author={Ivison, Hamish and Wang, Yizhong and Liu, Jiacheng and Wu, Zeqiu and Pyatkin, Valentina and Lambert, Nathan and Smith, Noah A and Choi, Yejin and Hajishirzi, Hannaneh},
  journal={Advances in neural information processing systems},
  volume={37},
  pages={36602--36633},
  year={2024}
}

\clearpage
\appendix

\section{Extended statistical analysis of perceptual parity}
\label{sec:app_stats}

To evaluate the subjective listening test results presented in Section 6.2, we transitioned from a naive paired hypothesis test to a Linear Mixed-Effects Model (LMM). The 660 user evaluations were presented individually and in randomized order to prevent anchoring bias. Because each participant rated both model outputs for each prompt, the dataset contains crossed random effects. 

We fit an LMM using the \texttt{statsmodels} library \cite{seabold2010statsmodels}, defining the subjective score (1--5 Likert scale) as the dependent variable and the model variation as the fixed effect. We included crossed random intercepts for both the \texttt{user\_id} and the \texttt{prompt} to explicitly model the variance introduced by individual rater harshness and the inherent difficulty of specific audio requests. The full regression results are provided in Table \ref{tab:lmm_summary}.

\begin{table}[h]
\centering
\caption{Linear Mixed-Effects Model regression results}
\label{tab:lmm_summary}
\begin{tabular}{lcccccc}
\toprule
\multicolumn{7}{l}{\textbf{Model:} MixedLM \hfill \hfill \textbf{Dependent Variable:} score} \\
\multicolumn{7}{l}{\textbf{No. Observations:} 660 \hfill \textbf{Method:} REML} \\
\multicolumn{7}{l}{\textbf{Log-Likelihood:} -905.9961 \hfill \textbf{Converged:} Yes} \\
\multicolumn{7}{l}{\textbf{Scale:} 0.7866} \\

\midrule
 & \textbf{Coef.} & \textbf{Std.Err.} & \textbf{z} & \textbf{P>|z|} & \textbf{[0.025} & \textbf{0.975]} \\
\midrule
\textbf{Intercept (Baseline)} & 2.921 & 0.190 & 15.364 & 0.000 & 2.549 & 3.294 \\
\textbf{Variation [RL Model]} & 0.115 & 0.069 & 1.668 & 0.095 & -0.020 & 0.250 \\
\midrule
\textbf{prompt Var} & 0.247 & 0.086 & & & & \\
\textbf{user\_id Var} & 0.281 & 0.150 & & & & \\
\bottomrule
\end{tabular}
\end{table}

The model converged successfully, yielding a non-significant main effect for the model variation ($p = 0.095$). The variance component estimates indicate that prompt-level variability ($\sigma^2 = 0.247$) was comparable in magnitude to participant-level variability ($\sigma^2 = 0.281$), validating our choice to use a crossed random-effects structure.

\subsection{Two One-Sided Tests (TOST) for Equivalence}

A non-significant $p$-value indicates a failure to reject the null hypothesis of no difference, but it does not formally prove that the two models are equivalent. To evaluate statistical equivalence, we conducted an equivalence test using the Two One-Sided Tests (TOST) framework \cite{lakens2017equivalence}.

We defined the margin of practical equivalence (Smallest Effect Size of Interest, or $\Delta$) as $\pm 0.25$ points on our 5-point Likert scale. The empirical standard deviation of the subjective ratings was approximately 0.88 (SD $\approx 0.88$). Under the common half-standard-deviation heuristic for practical significance, this corresponds to a margin of approximately $\pm 0.44$ rating points. Our chosen equivalence margin of $\Delta = 0.25$ is therefore more conservative.

In the TOST procedure at an $\alpha = 0.05$ significance level, equivalence is established if the $90\%$ Confidence Interval (CI) of the mean difference falls entirely within the $[-\Delta, +\Delta]$ bounds. Using the estimates for the Mean Difference ($+0.115$), Standard Error ($0.069$) and Critical value for the 90\% CI ($z_{0.95} = 1.645$), the $90\%$ Confidence Interval is calculated as:
\begin{equation}
    CI_{90\%} = 0.115 \pm (1.645 \times 0.069) =[+0.002, +0.229]
\end{equation}

The entire $90\%$ Confidence Interval $[+0.002, +0.229]$ falls within our equivalence bounds of $[-0.25, +0.25]$, we reject the presence of a practically meaningful difference ($p_{equivalence} < 0.05$). These results support that the perceptual quality of the 1.5B RL model is statistically equivalent to the GPT-4o mini baseline within the predefined equivalence margin of $\pm0.25$ rating points.

\section{Subjective evaluation user interface}

\begin{figure}
    \centering
    \begin{subfigure}{0.7\textwidth}
        \centering
        \includegraphics[width=\linewidth]{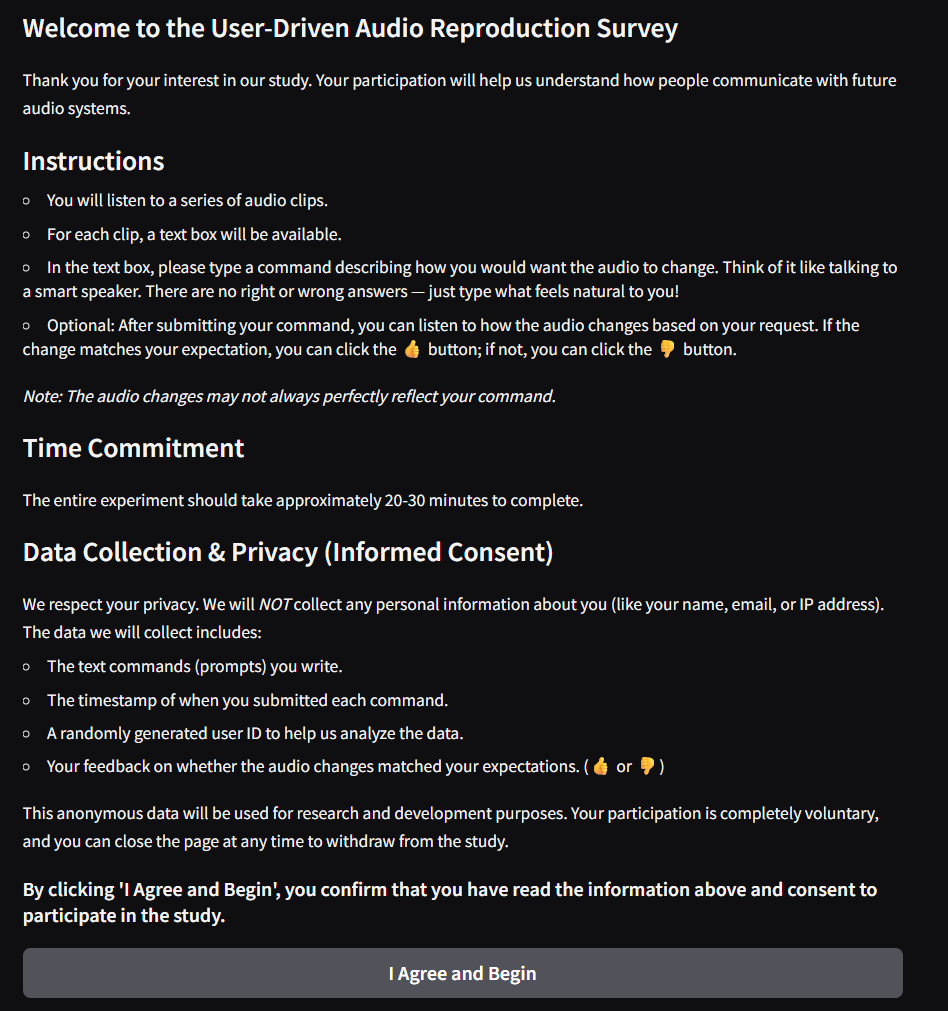}
        \caption{Introduction screen for the prompt elicitation phase, explaining the task to the user.}
        \label{fig:intro_exp1}
    \end{subfigure}
    \vspace{0.5cm}
    \begin{subfigure}{0.7\textwidth}
        \centering
        \includegraphics[width=\linewidth]{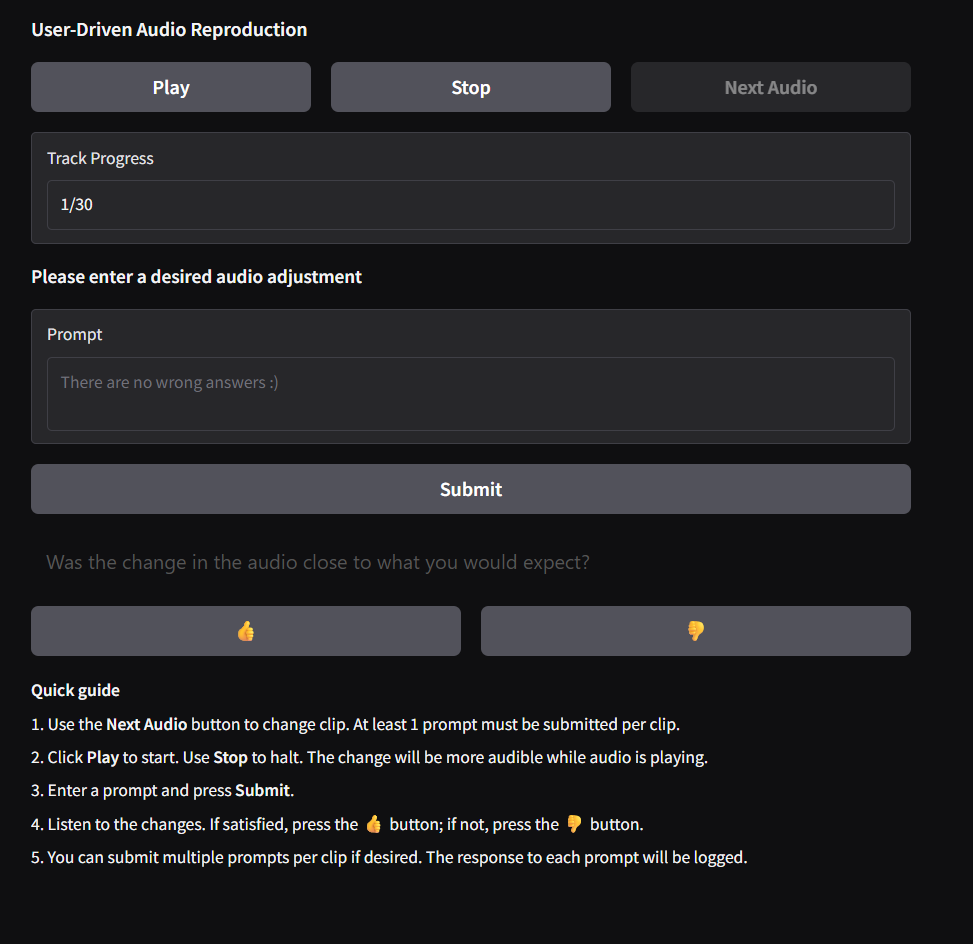}
        \caption{Main user interface for the prompt elicitation phase. Participants listened to an audio clip and could type any desired change into the text box.}
        \label{fig:exp_ui1}
    \end{subfigure}
    \caption{Screenshots from Phase 1 (Prompt Elicitation) of the subjective evaluation study.}
    \label{fig:appendix_ui_phase1}
\end{figure}

\begin{figure}
    \centering
    \begin{subfigure}{0.7\textwidth}
        \centering
        \includegraphics[width=\linewidth]{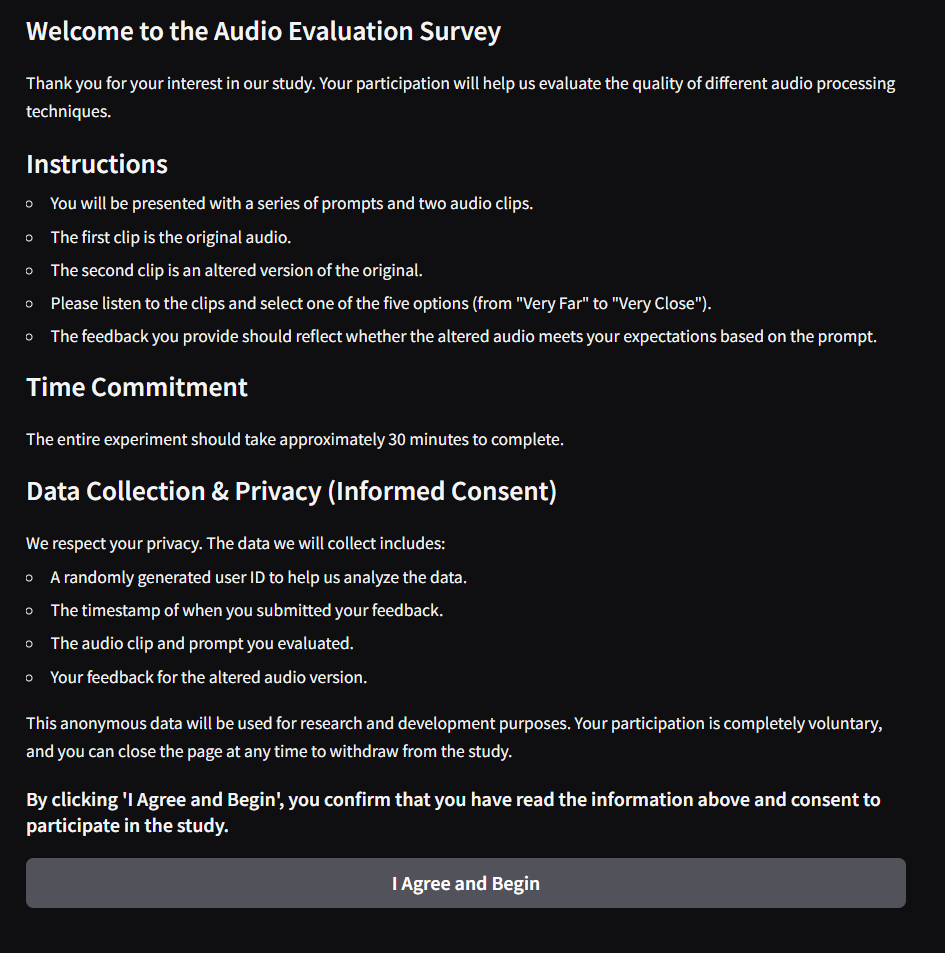}
        \caption{Introduction screen for the blind A/B comparison phase.}
        \label{fig:intro_exp2}
    \end{subfigure}
    \vspace{0.5cm}
    \begin{subfigure}{0.7\textwidth}
        \centering
        \includegraphics[width=\linewidth]{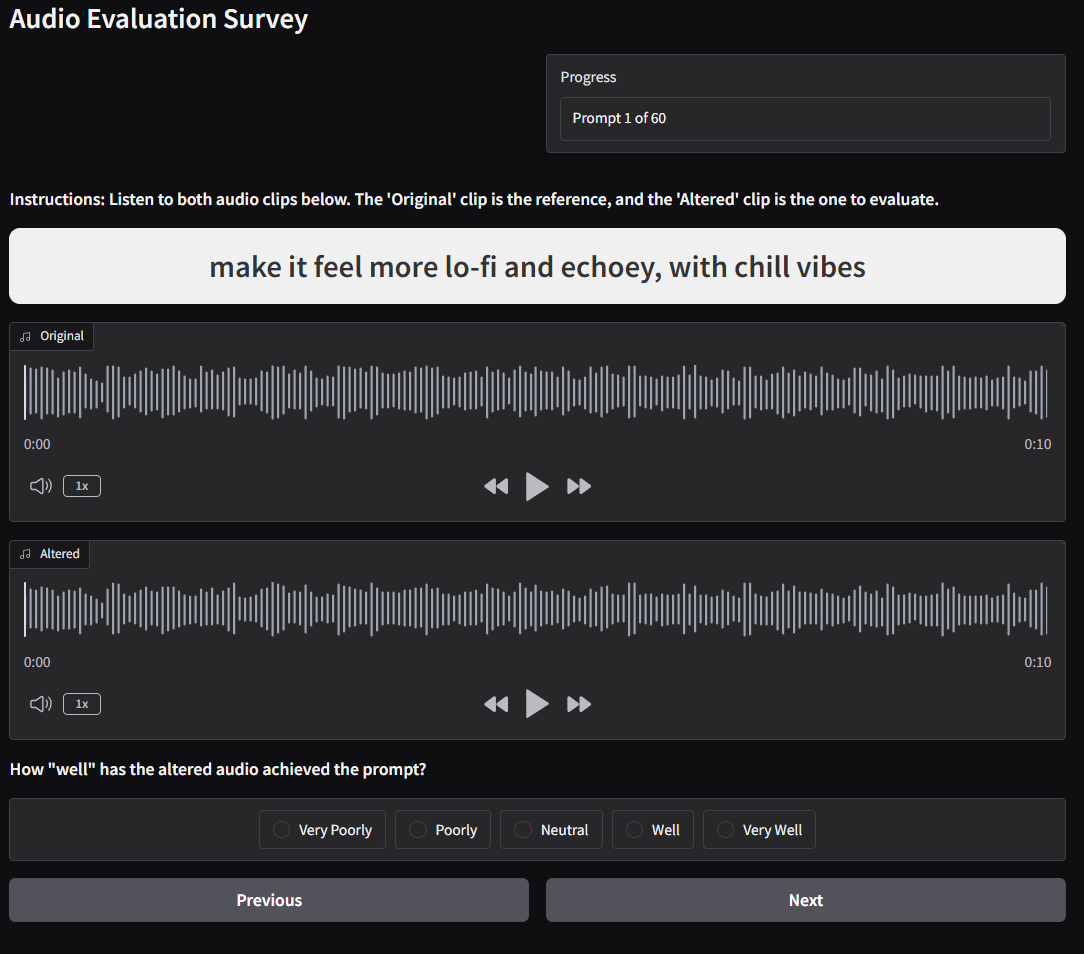}
        \caption{Main user interface for the A/B comparison. For each of the previously elicited prompts, users listened to two versions (``Original'' and ``Altered'') and rated them on a 5-point Likert scale.}
        \label{fig:exp_ui2}
    \end{subfigure}
    \caption{Screenshots from Phase 2 (Blind A/B Comparison) of the subjective evaluation study.}
    \label{fig:appendix_ui_phase2}
\end{figure}

To ensure the transparency of our subjective evaluation (Section \ref{sec:subjective}), we provide screenshots of the custom user interface presented to the participants. The study was conducted in two distinct phases: an initial unconstrained prompt elicitation phase (Figure \ref{fig:appendix_ui_phase1}) and a subsequent blind A/B model comparison phase (Figure \ref{fig:appendix_ui_phase2}).

\section{Ablation on KDE bandwidth selection}
\label{sec:appendix_kde}

To validate our choice of the bandwidth scaling factor for the Reflective-KDE ($h = 0.25 \cdot h_{\text{Scott}}$) introduced in Section \ref{subsec:rKDE}, we provide a visual ablation. Standard density estimators, such as Scott's Rule, typically assume a unimodal Gaussian distribution. While effective for simple datasets, applying an unscaled Scott's Rule to the highly multimodal and subjective preference data found in continuous audio control results in excessive smoothing. This smoothing destroys the local topological features representing specific user preferences.

\begin{figure}[h]
    \centering
    \includegraphics[width=\linewidth]{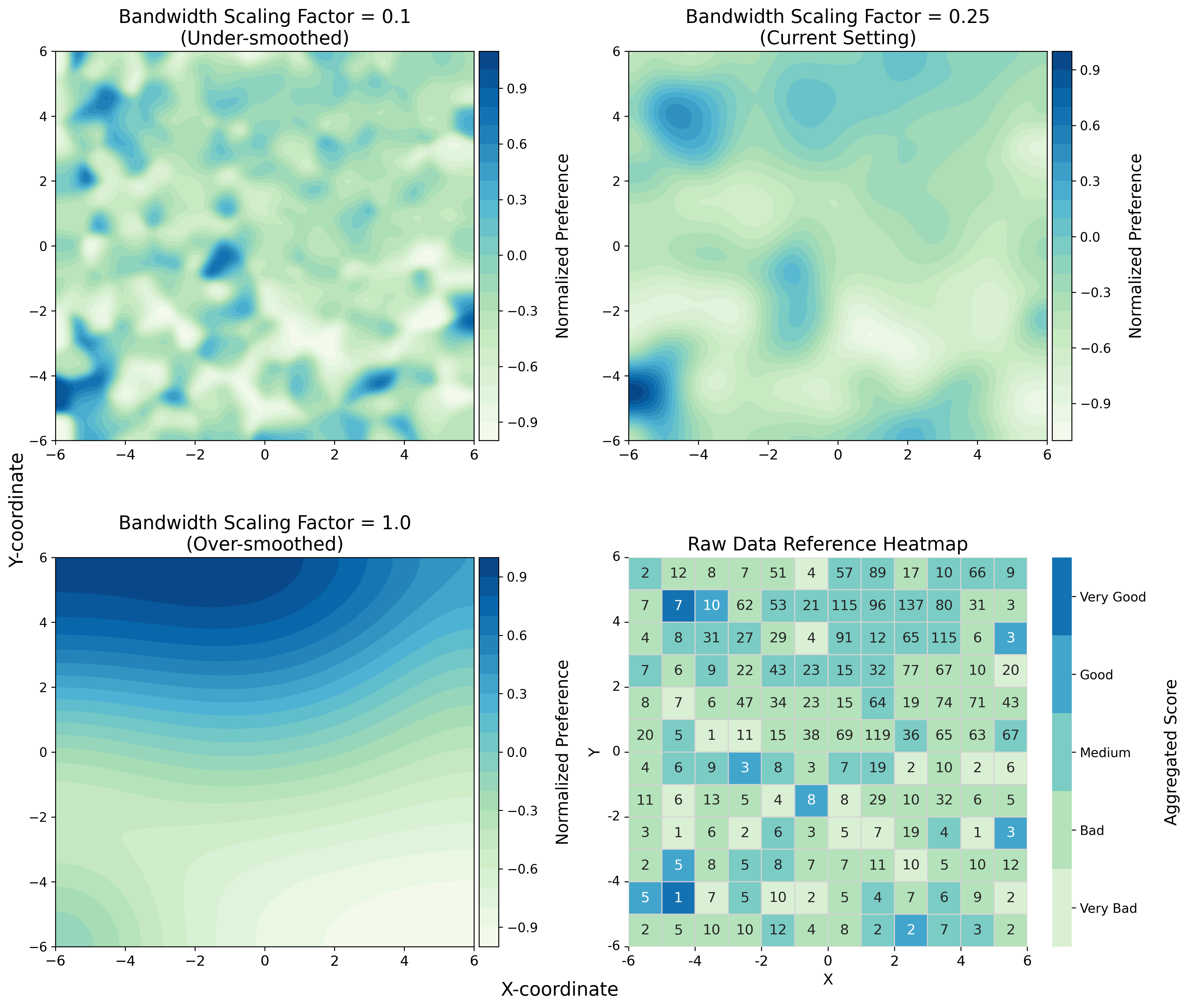}
    \caption{\textbf{Effect of KDE bandwidth scaling on the preference density.} The bottom-right panel displays a heatmap of the raw data for a representative prompt; cell colors indicate the aggregated preference score, while the integers inside the cells denote the number of user evaluations in that region. The remaining three panels show the resulting normalized reward surfaces using different bandwidth scaling factors. An under-smoothed factor (0.1, top-left) overfits to individual low-count data points, creating a noisy and fragmented reward signal. An over-smoothed factor (1.0, bottom-left) assumes unimodality, merging distinct preference modes into a single, uninformative gradient. Our chosen setting (0.25, top-right) provides an optimal balance, capturing the essential topological features and local maxima of the underlying preference manifold without overfitting to sparse noise.}
    \label{fig:kde_bw}
\end{figure}

Figure \ref{fig:kde_bw} demonstrates the effect of varying this scaling factor on the final non-parametric reward surface for a representative prompt. To contextualize the density estimation, the raw data heatmap (bottom-right) displays both the density of user evaluations (integers within cells) and the average preference score (cell color) across the discretized space. By scaling the bandwidth by $0.25$, we successfully counteract over-smoothing, interpolating high-confidence preference clusters while avoiding overfitting to isolated, low-sample noise.

\clearpage

\section{Experimental Details}

To facilitate full reproducibility of our alignment pipeline, Table \ref{tab:reproducibility} provides a comprehensive overview of the exact hyperparameters, hardware configurations, and methodological settings used throughout the training and evaluation phases.

\begin{table}[htpb]
\centering
\caption{Hyperparameters and Methodological Details for Reproducibility}
\label{tab:reproducibility}
\small
\begin{tabular}{@{}ll@{}}
\toprule
\textbf{Parameter} & \textbf{Value / Description} \\ \midrule
\multicolumn{2}{c}{\textbf{General \& Infrastructure}} \\ \midrule
Base Models & \texttt{Qwen/Qwen2.5-\{0.5B, 1.5B\}-Instruct} \\
Precision & \texttt{bfloat16} (Model weights and gradients) \\
Hardware & 1$\times$ NVIDIA L40 (48GB VRAM) \\
Distributed Training & None (Single GPU) \\
Optimizer & AdamW (\texttt{adamw\_torch}) \\
LR Scheduler & Linear decay (0 warmup steps) \\
Training Duration (GRPO) & $\approx$ 6 hours per model\\
Training Duration (DPO) & $\approx$ 1 hour per model\\ \midrule

\multicolumn{2}{c}{\textbf{Reward Surface \& Reflective-KDE}} \\ \midrule
Global Normalization Range & $[-1.0, 1.0]$ \\
Bandwidth Scaling ($h$) & $0.25 \cdot h_{Scott}$ \\
Numerical Stability Epsilon ($\epsilon$) & $10^{-7}$ \\
Malformed Format Penalty & $-1.0$ \\ \midrule

\multicolumn{2}{c}{\textbf{Stage 1: GRPO Training}} \\ \midrule
Learning Rate & $10^{-5}$ \\
Max Training Steps & $1000$ \\
Per-Device Train Batch Size & $4$ \\
Gradient Accumulation Steps & $32$ \\
KL Penalty ($\beta$) & $0.5$ (Optimal), Swept $\{0.0, 0.1, 0.5, 1.0\}$ \\
Number of Rollouts ($G$) & $16$ (Optimal), Swept $\{2, 4, 8, 16, 32\}$ \\ \midrule

\multicolumn{2}{c}{\textbf{Stage 2: Synthetic Data \& DPO Refinement}} \\ \midrule
Injected Peaks ($K$) & $10$ \\
Total DPO Pairs per Prompt & $20$ (10 Positive, 10 Negative) \\
Learning Rate & $5\times10^{-6}$ \\
Max Training Steps & $1000$ \\
Per-Device Train Batch Size & $1$ \\
Gradient Accumulation Steps & $12$ \\
KL Penalty ($\beta$) & $1.0$ (Optimal), Swept $\{0.0, 0.1, 0.5, 1.0\}$ \\ \midrule

\multicolumn{2}{c}{\textbf{Evaluation \& Inference}} \\ \midrule
Decoding Strategy & Greedy (\texttt{do\_sample=False, num\_beams=1}) \\
Max New Tokens & $32$ \\
\bottomrule
\end{tabular}
\end{table}

\section{Multi-seed experiments}
\label{sec:appendix_multiseed}
As noted in Section \ref{subsec:Hybrid_performance}, the performance of the hybrid pipeline fluctuates depending on the number of rollouts ($G$) used during the initial GRPO phase. To determine whether this stems from fundamental algorithmic instability or mere hyperparameter sensitivity, we evaluated our best configuration ($G=16$) across four independent training seeds. The model achieved a relatively consistent mean GMRR of $0.58 \pm 0.016$ with a 100\% parse rate.

\begin{table}[h]
\centering
\caption{Multi-seed performance of the optimal Qwen2.5-1.5B Hybrid model ($G=16$). The low standard deviation ($\pm 0.016$) demonstrates high training stability. Main model (GMRR = 0.60) not included.}
\label{tab:multiseed}
\small
\begin{tabular}{@{}l c c@{}}
\toprule
\textbf{Training Run} & \textbf{GMRR} & \textbf{Parse Rate} \\ \midrule
Seed 1 & 0.60 & 100\% \\
Seed 2 & 0.58 & 100\% \\
Seed 3 & 0.56 & 100\% \\
Seed 4 & 0.58 & 100\% \\ \midrule
\textbf{Mean ($\pm$ Std. Dev.)} & \textbf{0.58 $\pm$ 0.016} & \textbf{100\%} \\ \bottomrule
\end{tabular}
\end{table}


\end{document}